# Cryptanalysis on Four Two-Party Authentication Protocols

Yalin Chen
Institute of Information Systems and
Applications, NTHU, Tawain
d949702@oz.nthu.edu.tw

Jue-Sam Chou*
Dept. of Information Management
Nanhua University, Taiwan
jschou@mail.nhu.edu.tw
*: corresponding author

Chun-Hui Huang
Dept. of Information Management
Nanhua University, Taiwan
g6451519@mail.nhu.edu.tw

*Abstract¡* In this paper, we analyze four authentication protocols of Bindu et al., Goriparthi et al., Wang et al. and Holbl et al.. After investigation, we reveal several weaknesses of these schemes. First, Bindu et al.¡s protocol suffers from an insider impersonation attack if a malicious user obtains a lost smart card. Second, both Goriparthi et al.¡s and Wang et al.¡s protocols cannot withstand a DoS attack in the password change phase, i.e. an attacker can involve the phase to make user¡s password never be used in subsequent authentications. Third, Holbl et al.¡s protocol is vulnerable to an insider attack since a legal but malevolent user can deduce KGC¡s secret key.

## I. INTRODUCTION

Authentication protocols provide two entities to ensure that the counterparty is the intended one whom he attempts to communicate with over an insecure network. These protocols can be considered from three dimensions: type, efficiency and security.

In general, there are two types of authentication protocols, the password-based and the public-key based. In a password-based protocol, a user registers his account and password to a remote server. Later, he can access the remote server if he can prove his knowledge of the password. The server usually maintains a password or verification table but this will make the system easily subjected to a stolen-verifier attack. To address this problem, recent studies suggest an approach without any password or verification table in the server. Moreover, to enhance password protection, recent studies also introduce a tamper-resistant smart card in the user end. In a public key-based system, a user should register himself to a trust party, named KGC (Key Generation Center) to obtain his public key and corresponding private key. Then, they can be recognized by a network entity through his public key. To simplify the key management, an identity-based public-key cryptosystem is usually adopted, in which KGC issues user¡s ID as public key and computes corresponding private key for a user.

Considering computational efficiency in an authentication protocol, researchers employs low computational techniques such as secure one-way hash functions or symmetric key encryptions rather than much expensive computation like asymmetric key encryptions (i.e., RSA, ECC, ElGamal, and bilinear pairings). As considering communication efficiency, it usually to reduce the number of passes (rounds) of a protocol since the round efficiency is more significant than the computation efficiency.

The most important dimension of an authentication protocol is its security, and it should ensure secure communications for any two legal entities over an insecure network. Attackers easily eavesdrop, modify or intercept the communication messages on the open network. Hence, an authentication protocol should withstand various attacks, such as password guessing attack, replay attack, impersonation attack, insider attack, and man-in-the-middle attack.

In recent decade, many secure authentication protocols [1-41] were proposed. In 2008, Bindu et al. [14] proposed an improvement from Chien and Chen¡s work [3]. Their protocol is a smart-card based password authentication protocol and employs symmetric key cryptosystem. They claimed that their protocol is secure, provides user anonymity, and prevent from various attacks: replay attack, stolen-verifier attack, password guessing attack, insider attack, and man-in-the-middle attack. In 2009, Goriparthi et al. proposed a scheme [27] based on Das et al.¡s protocol [2] and can avoid the weakness existing in Chou et al.¡s [5]. Goriparthi et al.¡s protocol is also a smart card based password authentication protocol and bases on bilinear pairings. They claimed that their protocol is secure and can withstand replay attack and insider attack. In the same year, Wang et al. [31] also proposed an improvement based on Das et al.¡s protocol [2]. Their scheme is a smart card based password authentication protocol as well and uses secure one-way hash function. Also in 2009, Holbl et al. [40] improved from two identity-based authentication protocols, Hsieh et al. [1] and Tseng et al. [8]. Their protocols are neither password-based nor smart card based protocols. They employ identity-based ElGamal cryptosystem. Although all of the above schemes claimed that they are secure; however, in this paper, we will demonstrate some security vulnerabilities of these protocol in Bindu et al.¡s [14], Goriparthi et al.¡s [27], Wang et al.¡s [31], and Holbl et al.¡s work, correspondingly.

## II. REVIEW AND ATTACK ON BINDU ET AL.'S PROTOCOL

In this section, we first review Bindu et al.¡s protocol[14] and then show an insider attack launched by an insider who is supposed to have obtained another legal user¡s smart card.

### A. Review

There are three phases in Bindu et al.¡s protocol: the registration phase, the login phase, and the authentication phase.

In the registration phase, server S issues to user i a smart card which contains $m_i$ and $I_i$, where $m_i$=H($ID_i \oplus s$) $\oplus$ H($s$) $\oplus$ H($PW_i$), $I_i$=H($ID_i \oplus s$) $\oplus$ $s$, and $s$ is S¡s secret key.

When i wants to login to S, he starts the login phase and computes $r_i=g^x$ ($x$ is a random number chosen by i), $M=m_i \oplus$ H($PW_i$), $U=M \oplus r_i$, $R=I_i \oplus r_i$= H($ID_i \oplus s$) $\oplus$ $s \oplus r_i$, and $E_R[r_i, ID_i, T]$ ($T$ is a timestamp, and $E_R[r_i, ID_i, T]$ is a ciphertext encrypted by the secret key $R$). He then sends { $U$, $T$, $E_R[r_i, ID_i, T]$} to S.

In the authentication phase, after receiving { $U$, $T$, $E_R[r_i, ID_i, T]$} at time $T_s$, S computes $R= U \oplus$ H($s$) $\oplus$ $s$ =$M \oplus r_i \oplus$ H($s$) $\oplus$ $s$ =$m_i \oplus$ H($PW_i$) $\oplus r_i \oplus$ H($s$) $\oplus s$ = H($ID_i \oplus s$) $\oplus$ H($s$) $\oplus$ H($PW_i$) $\oplus$ H($PW_i$) $\oplus r_i \oplus$ H($s$) $\oplus s$ = H($ID_i \oplus s$) $\oplus r_i \oplus s$, decrypts $E_R[r_i, ID_i, T]$, checks to see if $T_s-T$ is less than $\Delta T$, and compares R with H($ID_i \oplus s$) $\oplus s \oplus r_i$ to see if they are equal. If they are, he sends {$T_s$, $E_R[r_s, r_i+1, T_s]$} to i, where $r_s=g^y$ and $y$ is a random number chosen by S. After that, i verifies the validity of the timestamp $T_s$, decrypts $E_R[r_s, r_i+1, T_s]$, and checks to see if $r_i+1$ is correct or not. If it is, S is authentic. Then, i sends {$E_{Kus}[r_s+1]$} to S, where $K_{us}=r_s^x=g^{xy}$. Finally, S decrypts the received message {$E_{Kus}[r_s+1]$} and checks to see if the value of $r_s+1$ is correct or not. If it is, i is authentic.

### B. Attack

If C lost his smart card and the card is got by an insider E, E can impersonate C to log into S. We show the attack in the following.

For that C¡s smart card stores $m_c$=H($ID_c \oplus s$) $\oplus$ H($s$) $\oplus$ H($PW_c$) and $I_c$=H($ID_c \oplus s$) $\oplus s$, and E¡s smart card stores $m_e$=H($ID_e \oplus s$) $\oplus$ H($s$) $\oplus$ H($PW_e$) and $I_e$=H($ID_e \oplus s$) $\oplus s$, suppose E gets C¡s smart card but doesn¡t have the knowledge of $PW_c$, E can choose a random number $x$ and computes $r_c=g^x$, $V= m_e \oplus I_e \oplus$ H($PW_e$)=H($s$) $\oplus s$, $M=I_c \oplus V=$ H($ID_c \oplus s$) $\oplus s \oplus$ H($s$) $\oplus s$ =H($ID_c \oplus s$) $\oplus$ H($s$) which equals $m_c \oplus$ H($PW_c$), $U=M \oplus r_c$, and $R= I_c \oplus r_c$. Then, E masquerades as C by sending { $U$, $T$, $E_R[r_c, ID_c, T]$} to S. After receiving the message, S computes $R=U \oplus$ H($s$) $\oplus s$ and compares $R$ with H($ID_c \oplus s$) $\oplus s \oplus r_c$. If they are equal, S sends C the message {$T_s$, $E_R[r_s, r_c+1, T_s]$}. E intercepts the message, decrypts $E_R[r_s, r_c+1, T_s]$, and uses $r_s$ to compute $K_{us}=r_s^x=g^{xy}$. E then can send a correct message {$E_{Kus}[r_s+1]$} to S, to let S authenticate him as C. In other words, insider E can successfully launch an insider attack if the user¡s smart card is lost.

More clarity, we demonstrate why $R=U \oplus$ H($s$) $\oplus s$ is equal to H($ID_c \oplus s$) $\oplus s \oplus r_c$ by the following equations.

$R=U \oplus$ H($s$) $\oplus s$
$= M \oplus r_c \oplus$ H($s$) $\oplus s$ ($\because U=M \oplus r_c$)
$= I_c \oplus V \oplus r_c \oplus$ H($s$) $\oplus s$ ($\because M=I_c \oplus V$)
$=$ H($ID_c \oplus s$) $\oplus s \oplus V \oplus r_c \oplus$ H($s$) $\oplus s$ ($\because I_c$=H($ID_c \oplus s$) $\oplus s$)
$=$ H($ID_c \oplus s$) $\oplus s \oplus$ H($s$) $\oplus s \oplus r_c \oplus$ H($s$) $\oplus s$ ($\because V$=H($s$) $\oplus s$)
$=$ H($ID_c \oplus s$) $\oplus s \oplus r_c$

## III. REVIEW AND ATTACK ON GORIPARTHI ET AL.'S PROTOCOL

In this section, we first review Goriparthi et al.¡s scheme [27] and then demonstrate a DoS attack on the password change phase of the protocol, which will make user¡s password never be used in subsequent authentications.

### A. Review

In the password change phase of Goriparthi et al.¡s protocol, when client C wants to change his password $PW$, he keys his $ID$ and $PW$ to his smart card. According their protocol, the smart card only checks $ID$ while no mechanism to verify the validity of $PW$. If the $ID$ is matched with the one stored in the smart card, the smart card will continuously ask C a new password $PW^*$, and then compute $Reg^*_{ID} = Reg_{ID}$ ¡ h($PW$) + h($PW^*$) = $s$¡h($ID$) + h($PW^*$), where $Reg_{ID} = s$¡h($ID$) + h($PW$) is issued by the server and stored in C¡s smart card in the registration phase, h(¡) is a map-to-point hash function, h:{0,1}*→$G_1$, and $G_1$ is a group on an elliptic curve. Finally, the smart card will replace $Reg_{ID}$ with $Reg^*_{ID}$.

### B. Attack

In the protocol, assume that an attacker temporarily gets C¡s smart card. He arbitrarily selects two passwords $PW'$ and $PW''$ as the old and the new ones, respectively. The smart card will then compute $Reg'_{ID} = Reg_{ID}$ ¡ h($PW'$) + h($PW''$) = $s$¡h($ID$) + h($PW$) ¡ h($PW'$) + h($PW''$) and replace $Reg_{ID}$ with $Reg'_{ID}$. This will make C¡s original password $PW$ never be used in subsequent authentications and thus cause denial of service.

## IV. REVIEW AND ATTACK ON THE PROTOCOL OF WANG ET AL.¡S PROTOCOL

In this section, we first review Wang et al.¡s protocol [31] and then show the protocol has the same weakness ¡ it suffers a DOS attack in password change phase ¡ like Goriparthi et al.¡s work [27].

### A. Review

In Wang et al.¡s protocol, C inserts his smart card, keys $PW$, and requests to change the password $PW$ to a new one $PW^*$. On receiving the request, the smart card computes $N_i^* = N_i \oplus$ H($PW$) $\oplus$ H($PW^*$) and replaces $N_i$ with $N_i^*$, where $N_i =$ H($PW_i$) $\oplus$ H($x$) is stored in C¡s smart card, $PW_i$ is chosen by the user when he registers himself to the remote server S, and $x$ is S¡s secret key..

### B. Attack

Obviously, this protocol also exits the same weakness like Goriparthi et al.¡s work [27]. Since if an attacker temporarily gets C¡s smart card, he can use two arbitrary values $PW'$ and $PW''$ to ask the smart card to update its storage through password change protocol. The smart card will compute $N_i' = N_i \oplus H(PW') \oplus H(PW'')$ and replace $N_i$ with $N_i'$. From then on, client C can never pass the subsequent authentications.

## V. REVIEW AND ATTACK ON THE PROTOCOL OF HOLBL ET AL.'S PROTOCOL

Holbl et al. [40] proposed two improvements of two-party key agreement and authentication protocols. In the following, we first briefly review their schemes and then present their weaknesses.

### A. Review of Holbl et al.¡s First Protocol

Holbl et al.¡s first protocol consists of three phases: the system setup phase, the private key extraction phase, and the key agreement phase.

In the system setup phase, KGC chooses a random number $x_s$ and keeps it secret. He computes $y_s = g^{x_s}$ as public key.

In the private key extraction phase, for each user who has identity $ID_i$, KGC selects a random number $k_i$, and calculates his private key $v_i = I_i k_i + x_s u_i \pmod{p¡1}$ and corresponding public key $u_i = g^{k_i} \pmod{p}$, where $I_i = H(ID_i)$.

In the key agreement phase, user A chooses a random number $r_a$, computes $t_a = g^{r_a}$, and then sends $\{u_a, t_a, ID_a\}$ to user B. After receiving $\{u_a, t_a, ID_a\}$, B chooses a random number $r_b$, calculates $t_b = g^{r_b}$, and then sends $\{u_b, t_b, ID_b\}$ back to A. Finally, A and B can respectively compute their common session key, $K_{AB} = (u_b^{I_b} y_s^{u_b} t_b)^{(v_a+r_a)} = g^{(v_b+r_b)(v_a+r_a)}$ and $K_{BA} = (u_a^{I_a} y_s^{u_a} t_a)^{(v_b+r_b)} = g^{(v_a+r_a)(v_b+r_b)}$, where $I_a = H(ID_a)$ and $I_b = H(ID_b)$.

### B. Attack on Holbl et al.¡s first protocol

Assume that an insider C calculates $I_c = H(ID_c)$ and $q = \gcd(I_c, u_c)$, and computes $w = I_c/q$, $z = u_c/q$, and $j = v_c/q$, where $v_c$ is C¡s private key. Hence, $\gcd(w, z) = 1$. Then, he can use the extended Euclid¡s algorithm to find $\alpha$ and $\beta$ both satisfying that $\alpha¡w + \beta¡z = 1$. As a result, he can obtain both $x_s$ and $k_c$, since $v_c = 1¡j_c¡q_c = (\alpha¡w + \beta¡z)¡j_c¡q_c = (\alpha¡I_c/q + \beta¡u_c/q)¡j¡q = (\alpha¡I_c + \beta¡u_c)¡j = I_c¡(\alpha¡j) + (\beta¡j)¡u_c$ and $v_c = I_c¡k_c + x_s¡u_c$, where $x_s$ is KGC¡s secret key and $k_c$ is a random number selected by KGC satisfying $u_c = g^{k_c}$. More clearly, the value $x_s$ he obtains is equal to $\beta¡j$.

After obtaining $x_s$, C can deduce any user¡s private key in the same manner. As an example, in the following, we demonstrate how C can deduces user i¡s private key, $k_i$. C calculates $I_i = H(ID_i)$ and $q_i = \gcd(I_i, u_i)$, computes $w_i = I_i/q_i$ and $z_i = u_i/q_i$, and then uses the extended Euclid¡s algorithm to compute $\gamma$ and $\varepsilon$ satisfying that $\gamma¡w_i + \varepsilon¡z_i = 1$. Finally, since $v_i = 1¡j_i¡q_i = (\gamma¡w_i + \varepsilon¡z_i)¡j_i¡q_i = (\gamma¡I_i/q_i + \varepsilon¡u_i/q_i)¡j_i¡q_i = (\gamma¡I_i + \varepsilon¡u_i)¡j_i = I_i¡(\gamma¡j_i) + (\varepsilon¡j_i)¡u_i$ and $v_i = I_i¡k_i + x_s¡u_i$, he can calculate $j_i = x_s/\varepsilon$ and thus obtains i¡s private key by computing $v_i = j_i¡q_i$. With the knowledge of i¡s private key, insider C can impersonate user i to communicate with any other legal user.

### C. Review of Holbl et al.¡s second protocol

Holbl et al.¡s second protocol consists of three phases: the system setup phase, the private key extraction phase, and the key agreement phase.

The system setup phase of this protocol is the same as the one in the first protocol.

In the private key extraction phase, with each user having his identity $ID$, KGC selects a random number $k_i$, and calculates i¡s private key $v_i = k_i + x_s¡H(ID_i, u_i)$ and public key $u_i = g^{k_i}$.

In the key agreement phase, user A chooses a random number $r_a$, computes $t_a = g^{r_a}$, and then sends $\{u_a, t_a, ID_a\}$ to user B. After receiving $\{u_a, t_a, ID_a\}$, B chooses a random number $r_b$, calculates $t_b = g^{r_b}$, and then sends $\{u_b, t_b, ID_b\}$ to A. Finally, A and B can compute their common session key, $K_{AB} = (u_b¡y_s^{H(ID_b, u_b)} \cdot t_b)^{(v_a+r_a)} = g^{(v_b+r_b)(v_a+r_a)}$ and $K_{BA} = (u_a¡y_s^{H(ID_a, u_a)} ¡t_a)^{(v_b+r_b)} = g^{(v_a+r_a)(v_b+r_b)}$, respectively.

### D. Attack on Holbl et al.¡s second protocol

Likewise, we can launch the same attack, as do in the first one, on this scheme. Since $\gcd(1, H(ID_c, u_c)) = 1$, an insider C can use the extended Euclid¡s algorithm to find $\alpha$ and $\beta$ both satisfying that $\alpha¡1 + \beta¡H(ID_c, u_c) = 1$. And since $v_c = k_c + x_s¡H(ID_c, u_c)$ and $1 = (k_c/v_c)¡1 + (x_s/v_c)¡H(ID_c, u_c)$, he can obtain both $x_s$ and $k_c$ by letting $x_s = \beta¡v_c$ and $k_c = \alpha¡v_c$, where $v_c$ is C¡s private key, $x_s$ is KGC¡s secret key and $k_c$ is a random number selected by KGC satisfying $u_c = g^{k_c}$. Consequently, similar to the result as shown in the attack of the first protocol, insider C can impersonate user i to communicate with any other legal user.

## VI. CONCLUSION

In the paper we have investigate four authentication protocols. In Bindu et al.¡s scheme [14], an insider can employ his own secrecy in the smart card issued from the server to successfully impersonate another user by getting the victim¡s smart card. In both Goriparthi et al.¡s and Wang et al.¡s schemes, their password change phases are easily subjected to a DOS attack, because no proper mechanism to verify user¡s input password. Finally, in Holbl et al.¡s scheme, any legal user can extract KGC¡s private key.

AUTHORS PROFILE

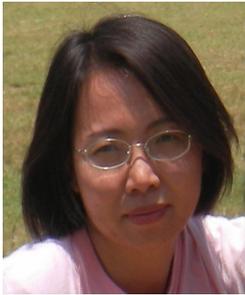

**Yalin Chen** received her bachelor degree in the depart. of computer science and information engineering from Tamkang Univ. in Taipei, Taiwan and her MBA degree in the department of information management from National Sun-Yat-Sen Univ. (NYSU) in Kaohsiung, Taiwan. She is now a Ph.D. candidate of the Institute of Info. Systems and Applications of National Tsing-Hua Univ.(NTHU) in Hsinchu, Taiwan. Her primary research interests are data security and privacy, protocol security, authentication, key agreement, electronic commerce, and wireless communication security.

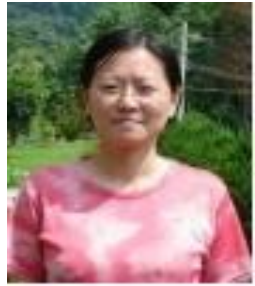

**Chun-Hui Huang** is now a graduate student at the department of Info. Management of Nanhua Univ. in Chiayi, Taiwan. She is also a teacher at Nantou County Shuang Long Elementary School in Nantou, Taiwan. Her primary interests are data security and privacy, protocol security, authentication, key agreement.

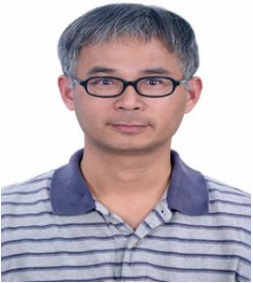

**Jue-Sam Chou** received his Ph.D. degree in the department of computer science and information engineering from National Chiao Tung Univ. (NCTU) in Hsinchu, Taiwan,ROC. He is an associate professor and teaches at the department of Info. Management of Nanhua Univ. in Chiayi, Taiwan. His primary research interests are electronic commerce, data security and privacy, protocol security, authentication, key agreement, cryptographic protocols, E-commerce protocols, and so on.